\documentclass[aps,prb,twocolumn,showpacs]{revtex4-1}
\usepackage{graphicx}
\usepackage{dcolumn}
\usepackage{bm}
\usepackage{amsmath}
\usepackage{tabularx}
\usepackage{float}
\usepackage{pifont}
\usepackage{rotating}  
\usepackage{hyperref}  
\usepackage{multirow}
\newfloat{widefig}{thp}{lop}
\begin{document}
\title{First-principles investigations into the thermodynamics of 
cation disorder 
and it's impact on electronic structure
and magnetic properties of spinel $Co\left(Cr_{1-x}Mn_{x} \right)_{2}O_{4}$}
\author{Debashish Das} %
\affiliation{Department of Physics, Indian Institute of Technology
Guwahati, Guwahati, Assam 781039, India} %

\author{ Subhradip Ghosh}
\affiliation{Department of Physics, Indian Institute of Technology
Guwahati, Guwahati, Assam 781039, India} %
\email{For correspondence: subhra@iitg.ernet.in}

\date{\today}

\begin{abstract}
Cation disorder over different crystallographic sites in spinel oxides is known
to affect their properties. Recent experiments on Mn doped multiferroic 
$CoCr_{2}O_{4}$ indicate that a possible distribution of Mn atoms among
tetrahedrally and octahedrally coordinated sites in the spinel lattice give 
rise to different variations in the structural parameters and saturation
magnetisations in different concentration regimes of Mn atoms substituting the
Cr. A composition dependent magnetic compensation behaviour points to the
role conversions of the magnetic constituents. In this work, we have 
investigated the thermodynamics of cation disorder in $Co\left(Cr_{1-x}Mn_{x}
\right)_{2}O_{4}$ system and it's consequences on the structural, electronic
and magnetic properties, using results from first-principles electronic
structure calculations. We have computed the variations in the cation-disorder
as a function of Mn concentration and the temperature and found that at the
annealing temperature of the experiment many of the systems exhibit cation
disorder. Our results support the interpretations of the 
experimental results regarding the qualitative variations in the sub-lattice
occupancies and the associated magnetisation behaviour, with
composition. We have analysed the variations
in structural, magnetic and electronic properties of this
system with variations in the compositions and the degree
of cation disorder from the variations in their electronic structures and by
using the ideas from crystal field theory. Our study provides a complete
microscopic picture of the effects that are responsible for composition dependent
behavioural differences of the properties of this system. This work lays down
a general framework, based upon results from first-principles calculations, 
to understand
and analyse the substitutional magnetic spinel oxides $A\left(B_{1-x}C_{x}
\right)_{2}O_{4}$ in presence of cation disorder.      
\end{abstract}

\maketitle
\section{Introduction}
Spinel oxides $AB_{2}O_{4}$ have been the subjects of continuously evolving
scientific research as they constitute a fascinating class of materials with
a plethora of interesting electrical, thermal, and magnetic properties. Above
these, spinel oxides are ideal materials for investigating 
structure-property relationships and consequent controlled engineering of
functional properties for device applications. The last traits are results of the
facts that many of them are known to exhibit substantial degrees of cation 
disorder \cite{dis1,dis2,dis3,dis4,dis5,dis6,dis7,dis8,dis9,dis10}
over the two types of sub-lattices, the tetrahedrally coordinated $A$
and the octahedrally coordinated $B$.The degree of cation disorder is 
represented by a single parameter $y$ which can take values between 0 and 1.
The "Normal spinel" compounds are those with $y=0$ when the $A$ atoms 
in $AB_{2}O_{4}$ compounds occupy the tetrahedral sites and $B$ atoms the 
octahedral sites. The "Inverse spinel"
compounds are the ones with $y=1$ where the tetrahedral sites are completely
occupied by the $B$ atoms, while the octahedral sites are occupied by equal
amounts of $A$ and $B$ atoms.
The degree of
cation disorder in a material depends on experimental conditions such as the 
annealing temperature \cite{anneal1,anneal2,anneal3} or the control of the
non-equilibrium disorder \cite{advfuncmat}. Various studies on a variety of
spinels have established that the electrical \cite{jpcm,advfuncmat}, electronic
\cite{mg} and thermal \cite{mg1,crystres} properties of these compounds can
be controlled by manipulating the degree of cation disorder. Thus, precise 
knowledge of the degree of cation disorder in a spinel compound is important
to understand it's fundamental properties.

Magnetic spinels with different magnetic constituent in $A$ and $B$ sub-lattices
have gained prominence as they widen the scope of functionalities of spinel 
oxides arising out of different magnetic interactions. Moreover, magnetism
offers greater perspectives regarding coupling between various degrees of freedom,
such as the lattice and the magnetic ones \cite{spphonon1,spphonon2}. Recently
discovered coupling between the charge and spin degrees of freedom,
the multiferroic property, in a few
chromite spinels,\cite{polarization,mncr2o4,cdcr2s4}
has generated considerable interests
regarding the understanding of the structure-property relationships and 
subsequent discovery of new multiferroic materials with improved functionalities.
Among these the oxide spinel $CoCr_2O_4$ is found to be a promising 
multiferroic material 
as it exhibits switchable electric polarisation
under reversal of magnetic field \cite{polarization}. The richness of it's
magnetic phase diagram comprising of various long range and short-range,
collinear and non-collinear magnetic structures \cite{cco1,cco2,cco3,cco4}
prompted researchers to examine the effects of substitution of one of the magnetic
atom with a new one on it's properties. Subsequently, new and interesting
phenomena like tunable Exchange Bias, temperature dependent magnetic
compensation and magneto-strictions were observed upon controlled substitution of Cr 
with Fe atoms \cite{padam-thesis,padam1,padam2,pss13}. These novel phenomena were
attributed to the composition dependent occupancy pattern of the substituted Fe
atoms in $CoCr_{2}O_{4}$ which violated the empirical idea of expected occupancy
pattern obtained from Mossbauer experiments \cite{moss} and also from the
theoretical model based upon the relative sizes of cations \cite{shannon}.

The interesting structure-property relationship upon substitution of Cr atoms
at $B$ sites in $CoCr_{2}O_{4}$ with Fe, led to a systematic investigation of Mn
substituted $CoCr_{2}O_{4}$ \cite{experiment}. The system $Co\left(Cr_{1-x}Mn_{x}
\right)_{2}O_{4}$ is expected to provide interesting insights into the
site preferences of the cations, and it's associations with the changes in
various properties as $x$ increases. 
This is due to the fact that while $CoCr_{2}O_{4}$ has a cubic ground state
(space group $Fd\bar{3}m$), $CoMn_{2}O_{4}$ is tetragonal with a significant
elongation along $z$ axis (space group $I4_{1}/amd$) \cite{anneal1}; there is no
evidence of any cation disorder in the former while non-negligible cation disorder
($y=0.22-0.39$)depending upon the annealing temperature is reported for 
the later \cite{anneal1,cmo}. Investigations into the structural and magnetic
properties \cite{experiment} revealed a deviation in the variations of the
lattice constant for smaller $x$ values ($x<0.15$) from the expected linear 
increase with increasing $x$, a structural distortion for $x>0.7$, a magnetic
compensation at $x \sim 0.25$, and three different patterns of variations in
magnetisations depending upon the ranges of $x$. The authors addressed this
non-regular behaviour in the lattice constants and the magnetisation along with
the phenomena of composition dependent magnetic compensation to the varying 
trends in the site preferences of the Mn atoms substituting the Cr. They 
explained the trends in the properties by assuming that the substituting Mn atoms
first occupy the tetrahedral $A$ sites upto the compensation point, and then
they first occupy one of the $B$ sites till a critical value of $x$ and finally
occupy the other $B$ site upon further increase in $x$. However, this investigation
neither quantified the cation disorder, nor provided any microscopic picture
behind the non-regular behaviours. 

With a motivation to understand the observed non-regular phenomena
in $Co\left(Cr_{1-x}Mn_{x} \right)_{2}O_{4}$ from microscopic point of view and
thus interpret them with a robust framework, in this work, we provide a detailed account
of the composition dependences and thermodynamics 
of the cation disorder in $Co\left(Cr_{1-x}Mn_{x} 
\right)_{2}O_{4}$ by combining Density Functional Theory (DFT) based first-principles
calculations and a thermodynamic model. These first-principles calculation based
results substantiate the experimental claim of the site preferences of the Mn
atoms by putting it on a solid theoretical ground. Concurrently, a systematic
study of the structural, magnetic and electronic properties as a function of
composition and degree of cation disorder is carried out. 
These results are interpreted in terms
of the elements in the crystal field theory and the electronic structures,
thus presenting the necessary microscopic picture governing the physics of this
system. In Section II, we present details of the models and the computational
techniques. The results are presented in Section III, followed by the conclusions.

\section{Details of calculations}
In an $AB_{2}O_{4}$ spinel, like $CoCr_{2}O_{4}$, with no cation disorder between
the sub-lattices, the tetrahedral sites are occupied by the $A$ atoms in +2 charge
state and the octahedral sites are occupied by the $B$ atoms in +3 charge state.
In case of cation disorder between the sub-lattices where degree of cation disorder
is denoted by the parameter $y$, the tetrahedral sites consist of a binary alloy
$A_{1-y}B_{y}$ and each of the octahedral sites has the composition 
$A_{y/2}B_{1-y/2}$. The degree of cation disorder $y$ at a finite temperature $T$
can be calculated from the thermodynamic consideration of cation distributions
by treating it as a simple chemical equilibrium \cite{navr}. Such modelling has been
proved to agree reasonably well with the experiments in most of the cases
\cite{advfuncmat,navr1,navr2,navr3,navr4,navr5,NN-app2}. In this model, the
configurational free energy of cation disorder per formula unit $\Delta F$
is given as
\[\Delta F= E_{c}- T\Delta S_{c} \]
$E_{c}$ is the cation disorder energy per formula unit, $T$ 
the temperature and
$\Delta S_{c}$ the configurational entropy which is calculated using the general
formula $\Delta S_{c}= -k_{B}\sum_{i,b} p_{i}^{b} \text{ln} p_{i}^{b}$; $p_{i}^{b}$ is
the concentration of the cation $b$ at the $i$-th sub-lattice. The equilibrium 
degree of cation disorder parameter $y_{0}$ at the given temperature $T$ can, then, be
obtained by minimising $\Delta F$ with respect to $y$.

The $Co\left(Cr_{1-x}Mn_{x}\right)_{2}O_{4}$ system, for a given value of $x$, 
can be represented as a $A\left(B_{1-x}C_{x} \right)_{2}O_{4}$ system.
Since we are dealing with a more complicated cation disorder as our system has
substitutional disorder with respect to one type of magnetic atom too, one 
has to
first decide on the compositions of each of the sub-lattices at a given value of
$x$ and for an arbitrary value of $y$. Ideally, the cation disorder should take
place between all three magnetic atoms. However, in our system, $Cr^{3+}$ has a
very strong preference to the octahedral sites; the octahedral site preference
energy(OSPE) of $Cr^{3+}$ has been found to be 46.7 kCal/mol while that of
$Mn^{3+}$ is 25.3 kCal/mol and that of $Mn^{2+}$ is 0 \cite{ospe}. Thus we can
assume that the cation disorder will take place only between the $A$ and the
$C$ atoms, that is, $Co$ and $Mn$ respectively. In spite of this simplification,
the compositions of the sub-lattices will be dependent upon the concentration $x$
of the $C$ atoms as shown in Table \ref{table0}. 
\begin{table}[bp]
    \caption{\label{table0} Compositions of tetrahedral and octahedral sub-lattices
in $A\left(B_{1-x}C_{x} \right)_{2}O_{4}$ for different ranges of $x$.}
\vspace{1 mm}
        \begin{tabular}{ccc}
        \hline\hline
Conc. & Composition at & Composition at \\ 
$x$        & $A$ site & $B$ site \\   \hline
$<0.5$    &  $A_{1-2xy}C_{2xy}$     & $B_{1-x}A_{xy}C_{x-xy}$            \\
$>0.5$    & $A_{1-y}C_{y}$ & $B_{1-x}A_{y/2}C_(x-y/2)$ \\
\hline \hline
\end{tabular}
\end{table}
For $x=0.5$, the two different patterns of sub-lattice 
occupancy shown in Table \ref{table0} will be identical. The reason behind
different compositions for different ranges of $x$ is due to the fact that the
total concentration at the tetrahedral sub-lattice cannot exceed 1; hence when
the $C$ content is greater than the $B$ content, the $C$ atoms, in excess of 1
must occupy the octahedral sub-lattice. In case of "complete
inversion" {\it i.e.} $y=1$, this is not satisfied if one uses the
site occupancy patterns for $x<0.5$,listed in Table \ref{table0}, 
for compositions with 
$x>0.5$.
The configurational entropy now assumes the form
\begin{eqnarray*}
\Delta S_c &=& -k_B [(1-2xy)\text{ln}(1-2yx)+ 2xy \text{ln}(2xy) \nonumber \\
    & &  \enskip \enskip \enskip \enskip +2(1-x)\text{ln}(1-x)+2xy \text{ln}(xy) \nonumber \\
    & &  \enskip \enskip \enskip \enskip  \enskip+2(x-xy)\text{ln}(x-xy)], \enskip \enskip \enskip \enskip 
    \enskip \enskip \enskip x<0.5 \nonumber \\
&=& -k_B [(1-y)\text{ln}(1-y)+y \text{ln}y +2(1-x)\text{ln}(1-x)\nonumber \\
    && +y \text{ln}(y/2)+(2x-y) \text{ln}((2x-y)/2)], \enskip \enskip x>0.5 
\end{eqnarray*} 
The non-configurational contributions towards the entropy, such as the
contributions from lattice vibrations, have been ignored in
the expression for $\Delta F$ as they have found to be typically small
compared to the configurational parts \cite{navr1,navr2,
NN-app2,advfuncmat}. However, for magnetic systems, contributions from the
magnetic entropy can be important in cases of severe fluctuations in 
magnetic moments of a particular constituent \cite{advfuncmat}. In the
present work, we have incorporated it in the expression for $\Delta F$ 
wherever appropriate, and will be discussed later.

The cation disorder energy $E_{c}$ is the energy difference between
a state with cation disorder $y$ and the "normal" state ($y=0$) and should
take into account the changes in the electrostatic and short-range interactions,
the crystal fields and charge states. Neil and Navrotsky \cite{NN-model} have
shown that $E_{c}$ follows a quadratic dependence on $y$ in the form
$E_{c}=\alpha y +\beta y^{2}$; $\alpha, \beta$ are the parameters. The
quadratic dependence was also empirically recognised by Kriessman and Harrison
\cite{harrison}. The advantage of this simple dependence of $E_{c}$ on
$y$ is that the total energies need to be calculated only for a few values of
$y$ and the energies can be fitted to the quadratic form so that the 
energy of cation disorder can be interpolated for any value of $y$ between $0$
and $1$. In the present work, we have calculated the total energies of our
systems, at each $x$, for three different $y$ values 0,0.5 and 1. The energies
are then fitted to the above quadratic form, in order to obtain the dependence
of $E_{c}$ on $y$ for the entire range of $y$. 

Modelling of the substitutional disorder in a condensed matter system is a
challenging problem. Although disorder of any arbitrary degree is now
routinely addressed by the mean field Coherent Potential Approximation (CPA)
\cite{cpa} in conjunction with multiple scattering Green's function methods
\cite{emto,kkr}, it has a serious limitation in dealing with systems 
having significant local distortions. In case of ternary spinels, the local
distortion is substantial even without cation disorder as is being seen by
the deviations of the oxygen parameter $u$ from it's ideal value of $0.25$
\cite{NN-app2,biplab,dd1}. With the introduction of a substituting element
having very different ionic radius, the distortion gets amplified \cite{dd2}.
Due to this, a recent investigation into the effects of Fe substitution in
$CoCr_{2}O_{4}$ using the CPA had limited itself to very low concentration
of Fe so that the local distortions do not have appreciable effects \cite{biplab}.
Since we are interested to explore the properties for the entire range of $x$,
varying between 0 and 1, we have taken the alternative approach of modelling the
substitutional disorder by constructing supercells of the spinel unit cell.
In this approach, addressing an arbitrary $x$ often becomes difficult as the
supercell size becomes prohibitively large. We have therefore considered 
$x=0.0625,0.125,0.25,0.5,0.75$ and $1$ for the present study. Such choices,
nevertheless, cover the composition regimes where the experimental observations
need to be addressed and interpreted.

For $Co\left(Cr_{1-x}Mn_{x} \right)_{2}O_{4}$ system, a unit cell of 14 atoms
can be considered. This unit cell is sufficient for simulating the 
three different
degrees of cation disorder given by $y=0,0.5,1$ for $x=0.5,0.75$ and $1$ only.
Therefore, we have considered a 56 atom supercell for all $x$ except $x=0.0625$;
a 112 atom supercell, the minimum one, was required for this concentration in
order to address all three degrees of cation disorder. 
In Table \ref{table01} we show the
configurations at tetrahedral and octahedral sites for each $x$ and $y$.
\begin{table}[bp]
    \caption{\label{table01} Compositions of tetrahedral and octahedral sub-lattices
in $Co\left(Cr_{1-x}Mn_{x} \right)_{2}O_{4}$ for different values of $x,y$ as
used in this work. The entries in the column "Cell size" denote the number
of atoms in the cell.}
\vspace{ 2 mm}
        \begin{tabular}{ccccc}
        \hline\hline
Conc. &Cell & $y$& Composition  & Composition  \\
$x$  & size &      & at $A$ site &  at $B$ site \\[0.5ex]    \hline
0.0625 & 112 & 0 & $Co_{16}$ & $Cr_{30}Mn_{2}$ \\
& &           0.5 & $Co_{15}Mn_{1}$ & $Cr_{30}Mn_{1}Co_{1}$ \\
& &           1.0 & $Co_{14}Mn_{2}$ & $Cr_{30}Co_{2}$ \\ \hline
0.125 & 56 & 0 & $Co_{8}$ & $Cr_{14}Mn_{2}$ \\
& &         0.5 & $Co_{7}Mn_{1}$ & $Cr_{14}Co_{1}Mn_{1}$ \\
& &         1.0 & $Co_{6}Mn_{2}$ & $Cr_{14}Co_{2}$ \\ \hline
0.25 & 56 & 0 & $Co_{8}$ & $Cr_{12}Mn_{4}$ \\
& &         0.5 & $Co_{6}Mn_{2}$ & $Cr_{12}Co_{2}Mn_{2}$ \\
& &         1.0 & $Co_{4}Mn_{4}$ & $Cr_{12}Co_{4}$ \\ \hline
0.50 & 56 & 0 & $Co_{8}$ & $Cr_{8}Mn_{8}$ \\
& &         0.5 & $Co_{4}Mn_{4}$ & $Cr_{8}Mn_{4}Co_{4}$ \\
& &         1.0 & $Mn_{8}$ & $Cr_{8}Co_{8}$ \\ \hline
0.75 & 56 & 0 & $Co_{8}$ & $Cr_{4}Mn_{12}$ \\
& &        0.5 & $Co_{4}Mn_{4}$ & $Cr_{4}Mn_{8}Co_{4}$ \\
& &        1.0 & $Co_{2}Mn_{6}$ &$Cr_{4}Mn_{6}Co_{6}$ \\ \hline
1.0 & 56 & 0 & $Co_{8}$ & $Mn_{16}$ \\
& &       0.5 & $Co_{4}Mn_{4}$ & $Mn_{12}$ \\
& &       1.0 & $Mn_{8}$ & $Co_{8}Mn_{8}$ \\ [1ex] 
\hline \hline
\end{tabular}
\end{table}  
The magnetic configuration in all cases have been taken to be Neel \cite{neel}
configuration, with
spins of $A$ and $B$ sub-lattices anti-aligned. Although the magnetic structure of
the end compounds of the system under investigation are non-collinear at 
low temperatures, it is
difficult to model, particularly since the progression of it with $x$ is not known. Thus we
have considered the collinear Ferrimagnetic structure which would help in qualitative
understanding of the magnetisation as the composition $x$ or degree of cation disorder $y$
is changed. Before fixing Neel configuration as the magnetic configuration, we have done
several calculations with different spin configurations at different sub-lattices. In almost all
cases, the Neel configuration came out to be energetically lowest. In cases where it was
not, the lowest energy spin configurations were lower by less than 0.1 meV per atom. This,
thus, further justifies consideration of Neel configuration for all $x$ and $y$. The small values
of energy differences between Neel and other configurations also indicate that the magnetic
structures may be spin disordered.

The total energies, the structural parameters, the magnetic moments and the electronic
structures were calculated by the DFT+U \cite{dft-u} method using Projector Augmented
Wave (PAW) \cite{paw} basis set as implemented in VASP \cite{vasp} code. The effects
of electron localisation were addressed by the approach of Dudarev {\it et al} \cite{dudarev}.
The Hund's coupling parameter $J$ was taken to be $1$ eV, while the Coulomb parameter
$U$ was taken to be $5$ eV for $Co$, $3$ eV for $Cr$ and $4$ eV for $Mn$. A plane wave cut-off
of $550$ eV and a $5 \times 5 \times 5$ mesh centred at $\Gamma$ point for Brillouin zone
integrations have been used throughout with the only exception for $x=0.0625$ where a 
$2 \times 2 \times 2$ mesh was enough to achieve an energy convergence of $10^{-7}$ eV.
Force convergences of $10^{-4}$ eV/$\AA$ were ensured during structural relaxations. 
 
\section{Results and Discussions}
\subsection{Temperature and concentration dependences of degree of cation disorder
$y$}
\begin{figure}[ht]
\includegraphics[width=8cm,height=8cm]{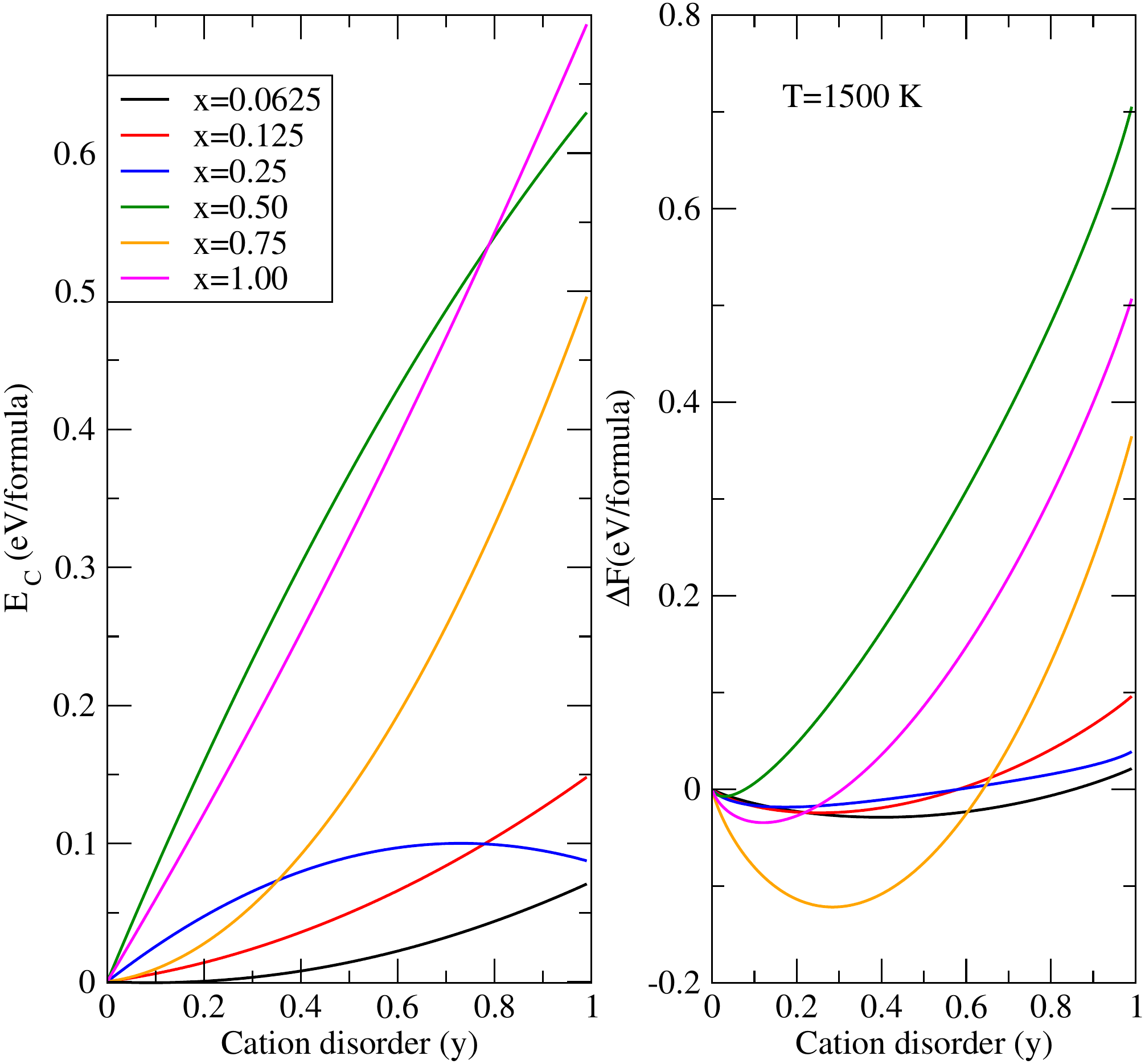}
\caption{\label{fig1} Variations of the cation disorder energy($E_c$)
(left panel) and the configurational free energy ($\Delta F$)(right panel)
 with degree of cation disorder $y$ of $Co\left(Cr_{1-x}Mn_{x} \right)_{2}O_{4}$, 
for different values of $x$, the Mn
concentration, at 1500 K, the annealing temperature of the experiment \cite{experiment}.
The equilibrium inversion parameter ($y_0$) at a given $T$ and
for a given $x$ is obtained from the minima of $\Delta F$.}
\end{figure}

\begin{figure}[ht]
\includegraphics[width=8cm,height=8cm]{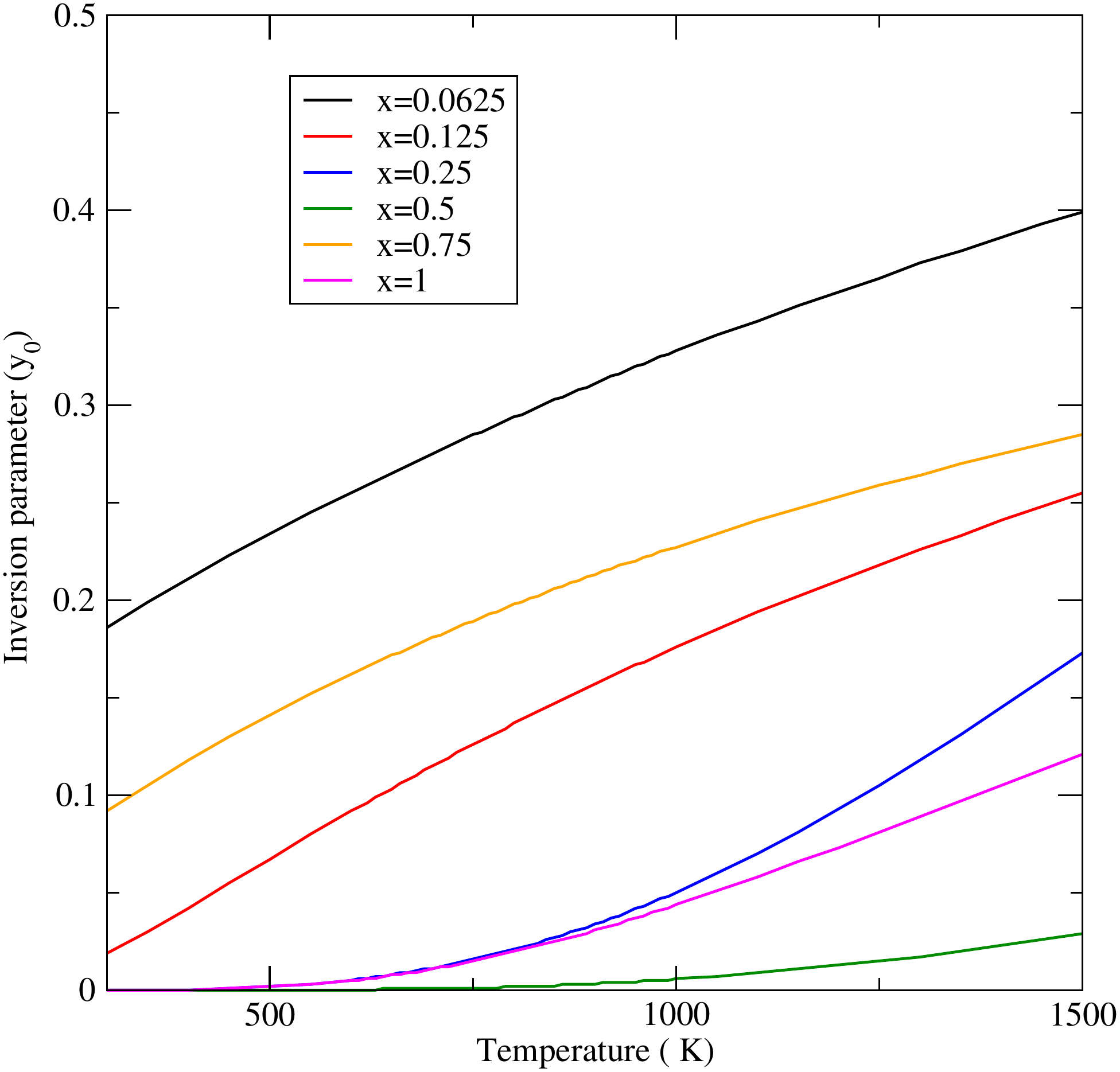}
\caption{\label{fig2} Temperature dependence  of the  
equilibrium inversion parameter ($y_0$) in $Co\left(Cr_{1-x}Mn_{x} \right)_{2}O_{4}$
for different $x$ for the temperature ranging from room temperature to the
annealing temperature of the experiment \cite{experiment}.}
\end{figure}

Before discussion of the physical properties of $Co\left(Cr_{1-x}Mn_{x} \right)_{2}O_{4}$,
it is required to find out the degrees of cation disorder between $Co$ and $Mn$ as 
a function of $Mn$ concentration $x$ and the temperature $T$. The temperature dependence
is important because earlier works on various spinel compounds have demonstrated that
$y$ may be quite sensitive to the annealing temperature \cite{anneal2,anneal3,navr5,NN-app2}.
In Fig. \ref{fig1} we show the dependences of the cation disorder energy $E_{c}$ and the
configurational free energy of cation disorder $\Delta F$ on $y$ for different $Mn$
concentration $x$, at a temperature $1500$ K which
is close to the annealing temperature of $1523$ K reported in the experiment
\cite{experiment}. We find that the cation disorder is zero for all values of $x$ if we do not
consider the configurational entropy. This implies that all substituting $Mn$ atoms will
be occupying the octahedral sites if the effect of the entropy is not included. Upon 
inclusion of the entropy term, states with cation disorder that is with non-zero values of
parameter $y$ are stabilised. The results suggest that the equilibrium value of $y$ at which
$\Delta F$ is a minimum goes towards $0$ as $x$ increases till 0.5, that is, up to the 
composition when the $Cr$ and $Mn$ contents in the system are identical. Further increase in
the $Mn$ content that is at $x=0.75$ increases the equilibrium value of $y$ before it is further
reduced when the $Mn$ substitution is complete at $x=1$. The quantitative variations of 
$y_{0}$, the equilibrium value of $y$, with temperature $T$, 
presented in Fig. \ref{fig2}, shows 
this qualitative behaviour clearly. 
At $1500$ K, $y_{0}=0.4, 0.26,0.17$ and nearly $0$ for
$x=0.0625,0.125,0.25$ and $0.5$ respectively. At $x=0.75$, 
the value of $y_{0}$ is $0.28$
which decreases to $0.13$ when $x=1$. This result, thus, suggests 
that in the beginning of
$Mn$ substitution in $CoCr_{2}O_{4}$, a significant amount of $Mn$ prefers to occupy the
tetrahedral positions, instead of the expected octahedral ones. As the $Mn$ content
increases, more and more $Mn$ occupies the octahedral positions until one reaches the point
where the $Mn$ content is equal to $Cr$ content ($x=0.5$). Further increase in $Mn$ content
initially puts some of it again in the tetrahedral sites, only to provide more preferences to the
octahedral sites as the content increases towards complete substitution of $Cr$. 
The importance of this result is that it supports the qualitative picture of site occupancies as
conjectured in the experiments \cite{experiment}. The differences with the picture provided
by the experimentalists is that they did not predict the re-emergence of the phenomenon of
$Mn$ atoms preference to tetrahedral sites after a critical composition. In the next sub-sections
we will provide an explanation of this.

The other important outcome of the thermodynamics of cation disorder is that the degree of
cation disorder in this system is not very robust as can be made out from the substantial variations
in $y_{0}$ with temperature. For any value of $x$, $y_{0}$ decreases continuously towards
$0$, and at room temperature the cation disorder for most of the compositions are 
insignificant. As cation disorder is suggested to be the reason behind non-regular behaviours
of magnetisation and phenomenon such as magnetic compensation \cite{experiment}, the 
importance of our results is that it provides the temperature range over which the degree of
cation disorder is appreciable and hence the experimental preparation of the samples should
be done accordingly. This also suggests that the cation disorder in these systems can be 
manipulated by controlling the temperatures. Such manipulations of cation disorder can 
substantially affect functional properties like the electrical conductivity as has been seen
elsewhere \cite{advfuncmat}.  
\subsection{Structural parameters and their variations with $x$ and $y$}
 \begin{table}[bp]
    \caption{\label{table1} Calculated cation-anion bond distances (in $\AA$) in $Co\left(Cr_{1-x}Mn_{x}
    \right)_{2}O_{4}$ for different $x$ and $y$ values.  }
    \vspace {2mm}
    \centering
\resizebox{\linewidth}{!}{
        \begin{tabular}{c|c|c|c|c|c|c} \hline \hline
         &       & \multicolumn{2}{|c|}{Tetrahedral site} & \multicolumn{3}{|c}{Octahedral site}       \\ \hline
x        & y     & Co-O     &  Mn-O    &    Cr-O  &  Mn-O       & Co-O       \\ \hline
0        & 0      & 2.00   &  -        &   2.01   &  -          & -            \\ \hline
         & 0      & 2.00   &  -        &   2.01   & 2.07   & -               \\
0.0625   & 0.5    & 2.00   &  2.04     &   2.01   & 1.98        & 2.10         \\
         & 1      & 2.00   &  2.04     &   2.01   &   -         & 2.10         \\ \hline  

         & 0      &  2.00  &  -        &   2.01   & 2.06   & -            \\
0.125    & 0.5    &  2.00  &  2.04     &   2.01   & 1.97        & 1.97         \\
         & 1      &  1.99  &  2.03     &   2.01   &  -          & 1.97             \\ \hline           

         & 0      &  2.00  &  -        &   2.01   &  2.07  & -            \\
0.25     & 0.5    &  2.00  &  2.04     &   2.01   &  2.07   & 1.97         \\
         & 1      &  1.99  &  2.03     &   2.01   &  -          & 1.96          \\ \hline 
         
         & 0      &  2.00   &  -       &   2.01   & 2.07   & -            \\
0.5      & 0.5    &  2.00  &  2.04     &   2.01   & 2.07   &  1.95            \\
         & 1      &  -     &  2.03     &   2.01   & -           &  1.95         \\ \hline           

         & 0      & 1.99   &  -        &   2.01   & 2.1 $\pm$ 0.15  & -            \\
0.75     & 0.5    & 2.00   & 2.04      &   2.01   & 2.06 $\pm$ 0.1 &  2.09 $\pm$ 0.03   \\
         & 1      & -      & 2.03      &   2.02  &2.1 $\pm$ 0.14   & 2.03 $\pm$ 0.04    \\ \hline           
                  
         & 0      & 1.99   &  -        &  -       & 2.13 $\pm$ 0.17  & -            \\
1.0      & 0.5    & 2.02   & 2.02 $\pm$ 0.04 &  -       & 2.1 $\pm$ 0.13 & 2.09 $\pm$ 0.07         \\
         & 1      &  -     & 2.02 $\pm$ 0.04 &  -       & 2.12 $\pm$ 0.14  & 2.04 $\pm$ 0.12  \\ \hline  \hline         
                       
         \end{tabular}}      
\end{table}
\begin{table}[bp]
    \caption{\label{table2} Calculated structural parameters of $Co\left(Cr_{1-x}Mn_{x}
    \right)_{2}O_{4}$ for different $x$ and $y$ values. The lattice constants $a$ and $c$ are in
    $\AA$. $u_{x},u_{z}$ are the oxygen parameters. }
    \vspace {2mm}
\begin{center}
\begin{tabular}
        { l@{\hspace{0.5cm}}  l@{\hspace{0.5cm}} l@{\hspace{0.5cm}} r@{\hspace{0.5cm}}c@{\hspace{0.5cm}}  c@{\hspace{1.0cm}} c@{\hspace{1.0cm}}c} \hline \hline
$x$        & $y$     & $c$    &  $a$   &    $u_z$ &  $u_x$       \\ \hline
0        & 0      & 8.43  &  8.43      &   0.262   &   0.262            \\ \hline
         & 0      & 8.44  &  8.44       &   0.262   & 0.262              \\
0.0625   & 0.5    & 8.44   &  8.44   &   0.262   & 0.262       \\
         & 1      & 8.45  &  8.45     &   0.262   &   0.262        \\ \hline  

         & 0      &  8.45  &  8.45      &   0.262  & 0.262           \\
0.125    & 0.5    &  8.45  &  8.45    &   0.262   & 0.262         \\
         & 1      &  8.43  &  8.43   &   0.262   &  0.262             \\ \hline           

         & 0      &  8.46 &  8.46        &   0.261  &  0.261            \\
0.25     & 0.5    &  8.43 &  8.43    &   0.262   &  0.262       \\
         & 1      &  8.42 &  8.42    &   0.263   &    0.262      \\ \hline 
         
         & 0      &  8.48  &  8.48     &   0.261 & 0.261           \\
0.5      & 0.5    &  8.44 &  8.44   &   0.263   & 0.263            \\
         & 1      &  8.40    &  8.40     &   0.265   & 0.265        \\ \hline           

         & 0      & 9.08   &  8.26      &   0.254   & 0.264          \\
0.75     & 0.5    & 8.68  & 8.35     &   0.259  & 0.263  \\
         & 1      & 8.67    & 8.42    &   0.262 &0.262  \\ \hline           
                  
         & 0      & 9.20   &  8.22      &  0.250     & 0.264          \\
1.0      & 0.5    & 9.06  & 8.24 &  0.257    & 0.264         \\
         & 1      &  8.70   & 8.45 &  0.256     & 0.262  \\ \hline  \hline         
                       
         \end{tabular}      
          \end{center}
\end{table} 
The first signature  of cation disorder upon $Mn$ substitution in $CoCr_{2}O_{4}$ 
was noted from the non-regular variations in their measured lattice constants with increasing $Mn$
content, in the low $Mn$ content regime \cite{experiment}. If the $Mn$, expected to be in
+3 charge state, had occupied the octahedral sites from the beginning, the lattice constant
should have increased linearly with $Mn$ content $x$ as
 $Mn$ atoms have larger ionic radii than $Cr$ at octahedral sites 
\cite{shannon}. We, therefore, first discuss the variations in the structural parameters
with concentration of $Mn$ as well as with variations in the degree of cation disorder.
In Table \ref{table1}, we present various cation-anion bond distances at
sites of different symmetries  and their variations with $x$ and $y$. In Table \ref{table2},
we present the lattice constants $a$ and $c$ along with the oxygen parameters
$u_x,u_z$ for various $x$ and $y$. The lattice constants and the
cation-anion bond distances are obtained from the DFT+U
calculations. The oxygen parameters are then obtained as \cite{ucalc}
\[u=\frac{-11+6r^{2}+\sqrt{33r^{2}-8}}{24\left(r^{2}-1 \right)} \] 
where $r=\bar{R}_{oct}/\bar{R}_{tet}$, the ratio of the average cation-anion bond
distances at the octahedral and the tetrahedral sites; $\bar{R}_{oct}, \bar{R}_{tet}$
are obtained by concentration averages of individual cation-anion bond distances.
The results show that for $x$ upto $0.5$, the systems retain
the cubic structure for all degrees of cation disorder, in agreement with the experiments.
For $x=0.75$ and $1$, a structural distortion leading to tetragonal phases are obtained,
again in agreement with the experimental observations. Quantitatively, the lattice constants
calculated by DFT+U are about $1-1.5 \%$ higher than the experimentally measured ones.
This is due to the choice of Generalised Gradient Approximation (GGA) 
\cite{gga} in describing the
Exchange-correlation part of the Hamiltonian. However, the qualitative agreement with 
the experiments is reasonably good, with a small increase in lattice constants with $x$
upto $0.5$. It is to be noted that the experimentally observed deviation in the 
lattice constant $a$ from the expected linear behaviour was for very small range
of $x$
($0< x < 0.25$) and the maximum change was only about $0.1 \%$. The maximum change
in the DFT+U calculated lattice constant for the range $x=0-0.25$ is about $0.3 \%$, if we
compare results of $y=0$ only. The comparisons of results with cation disorder $y=0$ only is justified
as we see that between $x=0-0.125$, the lattice constant $a$ does not change between 
$y=0$ and $y=0.5$ and that the equilibrium degree of inversion $y_{0}$ is well within this
range of $y$ moving towards $y=0$ as $x$ increases. For the concentrations $x=0.75,1$, we find that the
qualitative nature of variations in the lattice parameters ($c$ increases while $a$  decreases)
with $x$ is in agreement with the experiment. For $x=1$, the $c/a$ ratio (for $y=0$) obtained in our
calculations is $1.12$ which agrees very well with the experimental value of $1.09$
\cite{experiment}. We also find that the $c/a$ ratio decreases to $1.1$ when $y=0.5$. Since
our calculated equilibrium $y$ is in between $0$ and $0.5$, the agreement with experiment
would have been better had we considered the actual $y_{0}$ for the calculations.  For
$x=0.75$,  the calculated $c/a$ ratio varies from 1.1 to 1.04 as $y$ varies from $0$ to 
$0.5$ while the experimental value is $1.07$. Thus our results qualitatively agree with the
experiments as the calculated $y_{0}$ at this concentration is in between $y=0$ and $y=0.5$.
Comparison of our calculated $u$ parameters with experimental results for an arbitrary
$x$ is not possible due to unavailability of reported experimental values. The comparisons
can be made only for the end compounds. For $CoCr_{2}O_{4}$, the calculated $u$ value
of $0.262$ agrees well with the experimental value of $0.264$ \cite{cco2}. For
$CoMn_{2}O_{4}$, the calculated values of $0.255$ and $0.267$ for $u_x$ and $u_z$
($y=0$),
respectively, agree reasonably well with the experimental results of $0.230$ and $0.261$
\cite{anneal1}.
Thus, overall, our calculated structural parameters and their trends with $x$ are consistent
with the trends of calculated $y_{0}$ with variations in $x$.

We next turn our attentions to the variations of the lattice parameters with cation degree of
disorder $y$. The general trend seen in Table \ref{table2} is that with increase in $y$, the
lattice constants decrease, except at $x=0.0625$. This can be understood from the 
compositions of the tetrahedral and octahedral sites for different $y$ and $x$ and the variations
in their bond distances (Table \ref{table1}). For $x=0.0625$, we find that while the cation-anion
bond distances at tetrahedral sites are insensitive to the degree of cation disorder, the octahedral
$Mn-O$ bonds shorten considerably as $y$ changes from $0$ to $0.5$. On the other hand, larger $Co-O$ octahedral bond distances at $y=0.5$ state, 
bring the average octahedral 
cation-anion bond distances equal to that at $y=0$. Thus the lattice constants for $y=0$ and
$y=0.5$ are identical. For $y=1$, the octahedral sites are completely occupied by the $Co$
atoms making the average cation-anion bond distances associated with octahedral sites 
larger, resulting in a subsequent increase of the lattice constants with respect to $y=0,0.5$. 
The noticeable changes in the lattice constants with $y$ for $x=0.125,0.25$ and $0.5$
can be understood from the considerable decrease in the $Co-O$ octahedral bond distances
with cation disorder. The significant reduction in the $Co-O$ bond distances from their
values at $x=0.0625$ reduce the average octahedral cation-anion bond lengths and
consequently the lattice constants.  For $x=0.75$ and $1$, the tetragonal distortions bring in 
a lot of dispersions in the bond distances as can be seen from Table \ref{table1}. Once again,
it is the octahedral bond lengths which decide the trends in the lattice constants. The overall
decrease in the lattice parameter $c$ with $y$ is mainly brought about by 
the contracted $Mn-O$
bonds along the $z$-direction; the increase in $a$ is due to subsequent expansions of
the $Mn-O$ bonds in the $xy$-plane of the $Mn$ octahedra.  

We now try to provide explanations as to why the $Co-O$ octahedral bond distances
reduce dramatically for certain concentrations and degrees of cation disorder. It is 
expected that $Co$ at both sites will be having a +2 charge state. In that case, the effective
ionic radius of $Co$ should be $0.75 \AA$ at octahedral site and $0.58 \AA$ at
tetrahedral sites \cite{shannon}. The $Co-O$ bond distances at tetrahedral sites, thus, would be 
considerably lower than that for $Co$ at octahedral sites. Our calculations are 
consistent with this for $x=0.0625$ for all degrees of cation disorder. For $x=0.125,0.25$
and $0.5$, and for $y \neq 0$, our calculations show the opposite trend.  This indicates
that the $Co$ at octahedral sites for these cases would either be in a +2 charge state with
low spin or in a +3 charge state with either high or low spin(The effective ionic radii of
$Co^{2+}$ in low spin state, $Co^{3+}$ in high spin state and $Co^{3+}$ in low spin states
are $0.65 \AA, 0.61 \AA$ and $0.55 \AA$ \cite{shannon} respectively). From elementary crystal field theory \cite{cf}, it is known that the OSPE of a $d^{7}$ configuration is more than that of a $d^{6}$ configuration in high spin state,
while it is exactly opposite for a low spin state. Thus, the octahedral $Co$ atoms for these
$x$ and $y$ parameters are expected to be in a low spin +3 state ($d^{6}$ configuration).
The results on
magnetic moments, and the electronic structures, discussed in next sub-sections confirm
this and will be dealt in more detail.

The reason behind the tetragonal deformations at $x=0.75,1$ can also be understood from 
the crystal field theory.  In a comprehensive work, Dunitz and Orgel \cite{dunitz} had attributed
the electronic configurations at the octahedral and tetrahedral sites to the degrees of
distortion from cubic symmetry in spinels. Counting the average number of $t_{2g}$ and
$e_{g}$ electrons of tetrahedral and octahedral sites obtained from the charge states of
the cations, we find that for $y=0$, the octahedral sites have $(t_{2g})^{3}(e_{g})$ like
configurations, which will give rise to a large distortion in order to lift the degeneracy
associated with the $e_{g}$ states. With increase in the cation disorder, the presence of
$Co$ atoms in the octahedral site reduces the degeneracies associated with the $e_{g}$
orbitals. Thus, the degree of tetragonal distortions decrease as is seen from our calculated
results of decreasing $c/a$ with increasing $y$.  

Finally, we try to explain the reason behind re-appearance of non-negligible cation disorder
at $x=0.75$ after it had reduced to zero at $x=0.5$. The arguments are based upon the
bond distance results and elements of crystal field theory. From Table \ref{table1}, we find
that up to $x=0.5$, octahedral $Mn-O$ bond distances were constant for $y=0$ state.
At $x=0.75$, these bond lengths increase as a whole. From elementary crystal field theory,
we know that the crystal field parameter $\Delta$ at octahedral site is inversely 
proportional to the cation-anion distance  $R$ ($\Delta \sim R^{-5}$)\cite{cf}. Thus, when
the $Mn-O$ bonds increase in lengths at $x=0.75$, the corresponding octahedral crystal field 
will become weaker in comparison to that for other concentrations 
significantly. Consequently, the
octahedral crystal field stabilisation energy for the $Mn$ atoms will reduce as it is decided 
by the strength of $\Delta$, resulting in an increased preference of $Mn$ atoms towards
tetrahedral sites at $x=0.75$ in comparison to $x=0.5$. This is exactly reflected in the
re-emergence of non-zero $y_{0}$ at $x=0.75$. In the next sub-sections we provide more
conclusive evidences of the possible charge and spin states of various atoms in different
crystal sites which will corroborate the explanations given here. 
\subsection{Magnetic properties and their dependencies on $x$ and $y$} 
\begin{table*}[ht]
\footnotesize
    \caption{\label{table3} The inter-atomic nearest neighbour
    magnetic exchange interactions ($J_{ij}$ in meV;
    $i,j$ stand for tetrahedral ($T$) and octahedral ($O$) sites.) for various
    specie pair in $Co\left(Cr_{1-x}Mn_{x} \right)_{2}O_{4}$ with variations in $x$, the concentration
    of Mn. All calculations are done with $y=0$, the state with no cation disorder. The results for
    $y=0.5$ in case of  $x=0.5$ are given in parentheses. }
\begin{center}

       \begin{tabular}
          { l@{\hspace{0.5cm}} | l@{\hspace{0.5cm}} | l@{\hspace{0.5cm}} |l@{\hspace{0.5cm}} |l@{\hspace{0.5cm}}  |l@{\hspace{0.5cm}} |l@{\hspace{1.0cm}}} \hline \hline
       & \multicolumn{2}{|c|}{ $J_{TO}$} & \multicolumn{3}{|c|}{ $J_{OO}$} & $J_{TT}$\\ \hline
    $x$     &         $Co-Cr$ & $Co-Mn$  & $Cr-Cr$ & $Cr-Mn$  & $Mn-Mn$  & $Co-Co$                \\  \hline
0        &  -2.83        &   -           &-4.25        &  -              & -              & -0.49           \\           
0.0625   &  -2.83        &  -3.34        &-2.15         &-0.50            & -              & -0.50          \\ 
0.125    &  -2.83        &  -3.31        &-2.77         &-0.40            & -              & -0.50         \\ 
0.25     &  -2.80        &  -3.28        &-2.70         &-1.06            & -              & -0.47                    \\
0.50     & -2.59 (-2.36)       &  -3.38 (-3.42)       &-1.34(-3.88)         &-1.64 (-4.98)  & -3.39          & -0.42        \\ 
0.75     & -4.26        &  -2.91        & -            &-1.32            & -1.95 \text{(out of plane)}  & -0.31             \\  
            &                 &                  &               &                    & -8.64  \text{(in plane)}   & \\         
1        &  -            &  -3.47        &-             & -               & -1.054 \text{(out of plane)}   & -0.30    \\
          &               &                  &               &                  &  -9.46 \text{(in plane)} &     \\ \hline

 \end{tabular}
       \end{center}
\end{table*}
 \begin{table*}[ht]
    \caption{\label{table4} The total and atomic magnetic moments (in $\mu_{B}$ per formula unit)
    of $Co\left(Cr_{1-x}Mn_{x} \right)_{2}O_{4}$ for different concentrations $x$ of $Mn$ and
    for various degrees of cation disorder $y$ .}
\begin{tabular}
          { l@{\hspace{0.5cm}}  l@{\hspace{1.0cm}} l@{\hspace{0.5cm}} r@{\hspace{2.0cm}}c@{\hspace{1.0cm}}  c@{\hspace{1.0cm}} c@{\hspace{1.0cm}}c} \hline \hline

        & & Tetrahedral & site & Octahedral & site && \\ \hline
x        & y     & $\mu_{Co}$     & $\mu_{Mn}$    &    $\mu_{Cr}$  &  $\mu_{Mn}$    
                    & $\mu_{Co}$     & $\mu_{T}$   \\ \hline
0        & 0     & -2.68  & -      &  2.95  & -      & -      & 2.95           \\ \hline

         & 0      &-2.68  & -      &  2.95  &  3.86  & -      & 3.07             \\
0.0625   & 0.5    &-2.68  & -4.50  &  2.95  &  3.23  & 2.71   & 2.84           \\
         & 1      &-2.68  & -4.46  &  2.95  &  -     & 2.56   & 2.72         \\ \hline  

         & 0      & -2.68 &  -     & 2.95   & 3.87   &  -     & 3.19             \\
0.125    & 0.5    & -2.68 & -4.50  & 2.95   & 3.41   & 2.71   & 2.73           \\
         & 1      & -2.68 & -4.50  & 2.98   & -      & 0.18   & 1.76          \\ \hline  
         
         & 0      & -2.68 &  -     & 2.95   & 3.86   &  -     & 3.41              \\
0.25     & 0.5    & -2.67 &  -4.50 & 2.93   & 3.84   &  0.05  & 2.00           \\
         & 1      & -2.68 & -4.49  & 2.91   & -      &  0.02  & 0.58           \\ \hline           
         
         & 0      & -2.67  &  -     &  2.94  & 3.83   & -     &  3.89             \\
0.5      & 0.5    & -2.68  & -4.48  &  2.93  & 3.81   & 0.00  &  1.05             \\
         & 1      &  -     & -4.48  &  2.89  & -      & 0.10  & -1.81             \\ \hline
         
         & 0      & -2.68  &  -     &  2.92  & 3.85   & -     &  4.37            \\
0.75     & 0.5    & -2.69  & -4.49  &  2.88  & 3.80   & 2.72  &  2.53            \\
         & 1      & -      & -4.40  &  2.89  & 3.80   & 3.07  &  2.54          \\ \hline

         & 0      &  -2.71 & -     &   -     & 3.80   & -     & 4.84             \\
1        & 0.5    &  -2.68 & -4.48 &   -     & 3.70   & 2.71  & 3.01         \\
         & 1      &    -   & -4.39 &   -     & 3.78   & 3.07  & 3.02          \\ \hline \hline        

         \end{tabular}
\end{table*}
The magnetisation measurements \cite{experiment} on $Co\left(Cr_{1-x}Mn_{x} 
\right)_{2}O_{4}$ shows three distinct compositions ranges where the variations
of magnetisations with composition are different, although in all 
three regions, the
variations are linear. For $x \sim 0-0.25$, the magnetisation decreases and reaches 
the point of magnetic compensation. Further increase of $x$ shows a linearly increasing
magnetisation till another critical point $x \sim 0.65$ after which the magnetisation
decreases again with increase in $Mn$ content. The explanation of this 
behaviour was based upon the
canted spin structures observed in $CoCr_{2}O_{4}$ \cite{cco1} and $CoMn_{2}O_{4}$
\cite{anneal1}. The authors of Ref. \cite{experiment} considered that like $CoCr_{2}O_{4}$,
the canted spin structure will have opposing spin alignments at the two octahedral sites
with one of them (site $B_{1}$) aligning with the tetrahedral site. The magnetic
compensation behaviour was, thus, attributed to the initial occupation of $Mn$ atoms
at both tetrahedral and the other octahedral site $B_{2}$ with opposing spin alignments,
thus, cancelling the net moment as $Mn$ content is increased. The reason behind 
increase of the magnetisation after compensation was thought of due to $Mn$ atoms
occupying $B_{1}$ sites, diminishing the effects of $Mn$ atoms at $B_{2}$ sites gradually,
till the next critical point, after which the extra $Mn$ atoms start occupying the $B_{2}$
sites again, bringing a decreasing trend in magnetisation as $x$ increases. The 
experimentalists, however, did not substantiate their claim with detailed calculations. Their 
magnetisation measurements also indicated that the canting angles changed upon $Mn$
substitution.

Before interpreting the experimental results and checking the validity of the arguments
given in Ref. \cite{experiment} we first present results on nearest neighbour 
inter-atomic magnetic exchange interactions $J_{ij}$ for different
$x$ in Table \ref{table3}. These are calculated by mapping 
the DFT total energies for various collinear spin configurations on a Heisenberg 
Hamiltonian as is done elsewhere \cite{dd1}. The calculations
are done primarily for the "normal" spinel configuration ($y=0$).  Calculations for
other $y$ values require prohibitively large resources and often led to trouble in
convergences of self-consistent cycles. Hence, results for $y=0.5$ at $x=0.5$ is only
presented. The main point that can be made out of these results is that $J_{OT}$ and
$J_{OO}$ where $T$ stands for an atom at tetrahedral site and $O$ stands for an atom
at octahedral site, are comparable , $J_{TT}$ being negligible. The competing 
exchange interactions make  the spin structure non-collinear throughout the entire range
of $x$ as in agreement with experimental suggestion \cite{experiment}. 
For the compositions where the crystal
structure is cubic, $J_{OT}$ do not vary much while the two prominent $J_{OO}$
 compete as $x$ changes. As the $Cr$ content reduces, the $Cr-Mn$ exchange 
 interaction starts to strengthen at the expense of reduced $Cr-Cr$ interaction strength,
 until at $x=0.5$, the strengths of the two interactions become comparable. At $x=0.5$,
 $Mn$ atoms have other $Mn$ in their near neighbourhood due to their increasing
 content and thus have the strongest exchange interaction. The increased distances
 between $Cr$ atoms and between $Cr$ and $Mn$ atoms due to reduced $Cr$ content
 weaken $Cr-Cr$ and $Cr-Mn$ exchange interactions considerably. With further increase
 in the $Mn$ content, and with tetragonal distortion, $Cr$ has no $Cr$ or $Mn$ as nearest
 neighbours and the $J_{OO}$ is dominated by the $Mn$ atoms. A strong anisotropy
 in the exchange interactions is observed due to  the tetragonal distortions as was
 explained in Ref. \cite{dd2}. The spin structures at these compositions, are. therefore,
 expected to be more complicated.
 
 Since it is extremely difficult to model these non-collinear spin structures, particularly
 since no experimental information is available for $Mn$ substituted samples, we took
 recourse to the results obtained from calculations on collinear Neel structure in order to
 interpret the experimental results qualitatively. Table \ref{table4} presents the results
 on atomic and total magnetic moments for each $x$ and $y$. We find that the 
 magnetic moments of tetrahedral $Mn$ and
 tetrahedral $Co$ are insensitive to the changes in $x$ and $y$. As expected, the moment
 of tetrahedral $Co$ atom is close to $3 \mu_{B}$ implying that this $Co$ is in a 
 +2 charge state ($(e_{g})^{4}(t_{2g})^{3}$ configuration). However, the moment of $Mn$
 at tetrahedral site is rather close to $5 \mu_{B}$ which implies that this $Mn$ is 
 primarily in a 
 +2 charge state ($(e_{g})^{2}(t_{2g})^{3}$ configuration). The reason behind this charge 
 state of tetrahedral $Mn$ can be understood from the fact that the OSPE for $Mn^{2+}$
 in high spin state is 0 while that of $Mn^{3+}$ is $25.3$ kCal/mol \cite{ospe} and thus 
 the probability of assuming +2  charge state is greater for tetrahedral $Mn$ in high spin state.
 Significant variations in the magnetic moments with changes in $Mn$ composition and 
 in degree of cation disorder is observed for $Co$ atoms at the octahedral sites, while the
 moments of $Cr$ and $Mn$ at octahedral sites have insignificant variations. The
 $Cr$ and $Mn$ atoms at octahedral sites are in high spin, +3 charge states for all $x$ 
 and $y$. The much greater OSPE of high spin $Mn^{3+}$ and $Cr^{3+}$ states in
 comparison to $Mn^{2+}$ are responsible for this. The $Co$ atoms at the octahedral 
 positions oscillate between high spin and low spin states depending upon $x$ and $y$.
 For $x=0.0625$, irrespective of the degree of cation disorder,
 the tetrahedral $Co$ are in high spin and +2 charge states as is reflected
in their
 moments being $\sim 3 \mu_{B}$. For $x=0.125$, while $50 \%$ of cation disorder ($y=0.5$)
 still keeps $Co$ spin and charge states same, complete "inversion" ($y=1$) quenches the
 $Co$ moment leading to a low spin state with moment $\sim 0$. This trend continues till
 $x=0.5$ and is independent of the degree of cation disorder. The high spin state is regained
 at $x=0.75$ and remains intact for $x=1$ with $\mu_{Co} \sim 3 \mu_{B}$ 
 irrespective of the degree of cation disorder. As was argues in the previous sub-section,
 the relative OSPE values \cite{cf}for $Co^{2+}$ and $Co^{3+}$ clearly demonstrate 
 that in the low spin state, $Co^{3+}$ would have a greater preference towards octahedral
 sites. One can relate the octahedral $Co-O$ bond distances (Table \ref{table1}) with the
 low spin +3 charge state in this case using elementary crystal field theory. The 
 significant decrease in octahedral $Co-O$ bond distances for a range of $x$ and $y$,
 increases the octahedral $Co$ crystal field. As a result these $Co$ pair up the electrons.
 As the magnetic moment is close to zero, the configuration must be $(t_{2g})^{6}(e_{g})^{0}$
 which is consistent with the preferred +3 charge state as predicted from results on OSPE.
 The regaining of the high spin state at $x=0.75$ can be understood from the increase in
 $Co-O$ octahedral bond lengths due to the structural relaxations, and subsequent weakening
 of the octahedral $Co$ crystal field. As was mentioned earlier, due to the low spin state of octahedral $Co$  while the rest of the atoms at different sites are in high spin states,
 an additional magnetic entropy of $2{k}_{B}xy \text{ln} 2$ is added to $\Delta F$ for the
 select $x$ and $y$ values. A non-zero $y_{0}$ value for these concentrations are obtained 
 only if this magnetic entropy term is added. Thus, a cation disordered state at this
 concentrations is driven by the changes in the magnetic entropy.
 
 Although the results are for collinear spin arrangements, in conjunction with
 the calculated $y_{0}$ values for each $x$, a qualitative variation in the total
 magnetic moment quite similar to the experimentally observed can be extracted
 by careful analysis of the results. Since 
 $y_{0}$ for $x=0.0625$ is $\sim 0.4$, the expected magnetic moment would be close to
 $2.84 \mu_{B}$, the calculated value for $y=0.5$. As the $y$ is in the range of $0.17-0.4$
 for $x$ upto 0.25, the expected values of magnetic moments may show
a decreasing trend as $x$
 increases. An increase in the moment upon further increase of $x$, that is in the range
 $0.25 <x <0.75$, is expected as the $y_{0}$ continuously reduces making the cation
 disorder close to zero, and thus the appropriate moments to look at from Table
 \ref{table4} would be the ones with $y=0$ which shows continuous increase with $x$.
 Our results in the range $x=0.75-1$, however, do not quite seem to follow experimental
 behaviour qualitatively. At $x=0.75$, the magnetic moment when $y=0.28$, the $y_{0}$
 value at this concentration, should be close to $3.5 \mu_{B}$ if the results between $y=1$
 and $y=0.5$ are interpolated. This will indicate a decrease of moment as $x$ changes
 from $0.5$ to $0.75$ agreeing with the experimental behaviour. But the same 
 interpolation in case of $x=1$ puts the magnetic moment at this concentration close to
 $4.5 \mu_{B}$ which means an increase with respect to result at $x=0.75$, in contradiction
 to the experimental trend. In spite of this disagreement, it is interesting that the
 results from collinear spin arrangement follow the experimental trend for a significant range
 of composition.  
 
 A qualitative model explaining the trends in the experimentally measured magnetisation
 can now be constructed by using the calculated atomic moments, the site occupancy
 patterns given in Table \ref{table0} and the alignments of sub-lattice spins obtained 
 in the experiments. The experimental spin structure makes a distinction between two
 octahedral sub-lattices which our collinear spin arrangement does not. If instead of
 considering the occupancies of both octahedral sub-lattices to be identical, 
we consider that after part of $Mn$ occupying the tetrahedral sites 
in accordance with the pattern depicted in Table \ref{table0}, the rest of 
the $Mn$ for a given $x$ 
 completely occupies the $B_{2}$ sub-lattice which is anti-aligning with the other octahedral
 $B_{1}$ and the tetrahedral sub-lattice, then the experimental trends of magnetisation
 for $x=0-0.25$ is qualitatively reproduced. This can be understood the following way:
 If we consider the DFT calculated atomic magnetic moments as the moments of 
 individual atoms in this picture and consider that the $B_{2}$ sub-lattice has $Cr$ and $Mn$
 while $B_{1}$ has $Co$ and $Cr$, and the $y$ value for each $x$ is taken to be equal to that of the calculated $y_{0}$, then with increase in $x$, 
the contribution of $B_{2}$ sub-lattice will
 gradually increase in comparison to the other two which will reduce the net magnetic
 moment steadily. The increasing dominance of $B_{2}$ will be due to the fact that 
 with increase in $x$, the $y$ steadily decreases and with rather small $x$, the content of
 $Mn$ at the tetrahedral site is always small. Since octahedral $Cr$ and tetrahedral $Co$ 
 have identical moments, their moments from $B_{2}$ and the tetrahedral sites will nearly
 cancel each other. $Mn$ being the carrier of larger magnetic moments would have the 
 maximum effect. Thus the $Mn$ at $B_{2}$ site would control the magnetisation variation
 in this concentration range. With further increase in $x$, the $y$ value is supposed to
 be getting smaller and nearly vanish for $x=0.5$. Thus, in the concentration range $x=0.25-
 0.5$, the $Mn$ is going to occupy the octahedral sites mostly. If we now once again let
 the occupancies of the two octahedral sites be different with most of the extra $Mn$ occupying
 the $B_{1}$ site, then the magnetic moment of the system will increase as $x$ increases.
 This is because of the fact that since the occupancy of $B_{2}$ sub-lattice would not
 change much from what it was for $x=0-0.25$, more $Mn$ content at $B_{1}$ site
 with their spins anti-aligning to those of $Mn$ in $B_{2}$ site would increase the
 total moment as the spins of $B_{1}$ and the tetrahedral sites align. This occupancy
 pattern would continue till $x=0.5$ and possibly a little further up till about the critical
 point ($x \sim 0.7$) observed experimentally. Since $y=0$ when $x=0.5$, the expected
 occupancies in tetrahedral sub-lattice will be $Co$ only, while both 
 octahedral sub-lattices will have equal amounts of $Cr$ and $Mn$. With
 further increase in $x$ and with expected $y$ value nearly zero, $B_{2}$ 
will have equal amounts
 of $Cr$ and $Mn$ and $B_{1}$ will be $Mn$ rich $Mn-Cr$ alloy, thus 
 shooting up the magnetic moment further. At $x=0.75$, the cation disorder returns.
 Now the cation disorder will be between tetrahedral and $B_{1}$ sites, with $Co$
 occupying the $B_{1}$ sites but the extra $Mn$ mostly occupies the $B_{2}$ sites.
 As a result, the total moment steadily decreases as $x$ increases. This picture, thus, 
 not only supports what was espoused by the experimentalists, but also puts 
 it on a solid theoretical footing by using information from first-principles calculations.            

 \subsection{ Electronic structures and their variations with $x$ and $y$}
 
 \begin{table}[bp]
    \caption{\label{table5} The calculated Band gaps (in eV) of $Co\left(Cr_{1-x}Mn_{x} \right)_{2}
    O_{4}$ for different $Mn$ concentrations $x$ and different degrees of cation disorder $y$.}
\vspace{0.5mm}
\centering
        \begin{tabular}{c|ccl} \hline \hline
        & \multicolumn{3}{l}{Cation disorder $y$}  \\ \hline
x        &  0.0  & 0.5  & 1.0     \\ \hline
0.0    &  2.1     &   -       & -           \\        
0.0625   &  0.77     &  0.66    &  1.99      \\ 
0.125    &  0.70     &  0.33    &  1.61      \\ 
0.25    &  0.38     &  0.64    &  1.53      \\
0.50    &  0.15     &  0.58    &  1.44      \\ 
0.75    &  0.23     &  0.48    &  0.62      \\ 
1.0    &  0.33     &  0.00    &  0.56      \\ \hline \hline
            \end{tabular}
\end{table}

\begin{figure}[ht]
\includegraphics[width=8.0cm,height=10.0cm]{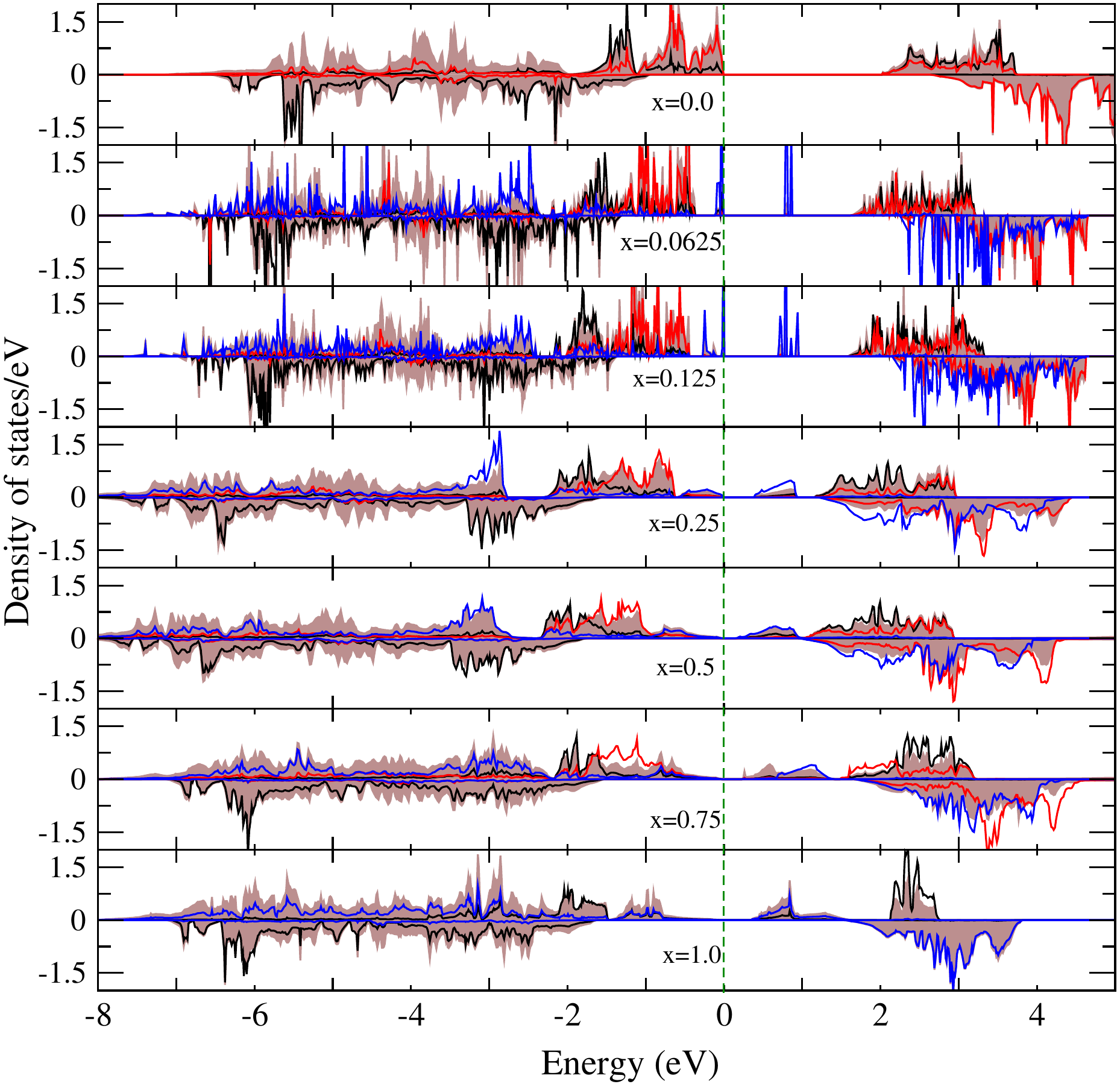}
\caption{\label{fig3} Total and atom-projected densities of states of 
$Co\left(Cr_{1-x}Mn_{x} \right)_{2}O_{4}$
for different $x$. The results are for zero cation disorder ($y=0$). The 
results for $CoCr_{2}O_{4}$ ($x=0$) are also included. Here 
 the total densities of states are denoted by brown shades. 
The black, red and blue
curves represent atom projected densities of states of Co at tetrahedral
sites, Cr and Mn at octahedral sites, respectively.}
\end{figure}

\begin{figure}[ht]
\includegraphics[width=8.0cm,height=10.0cm]{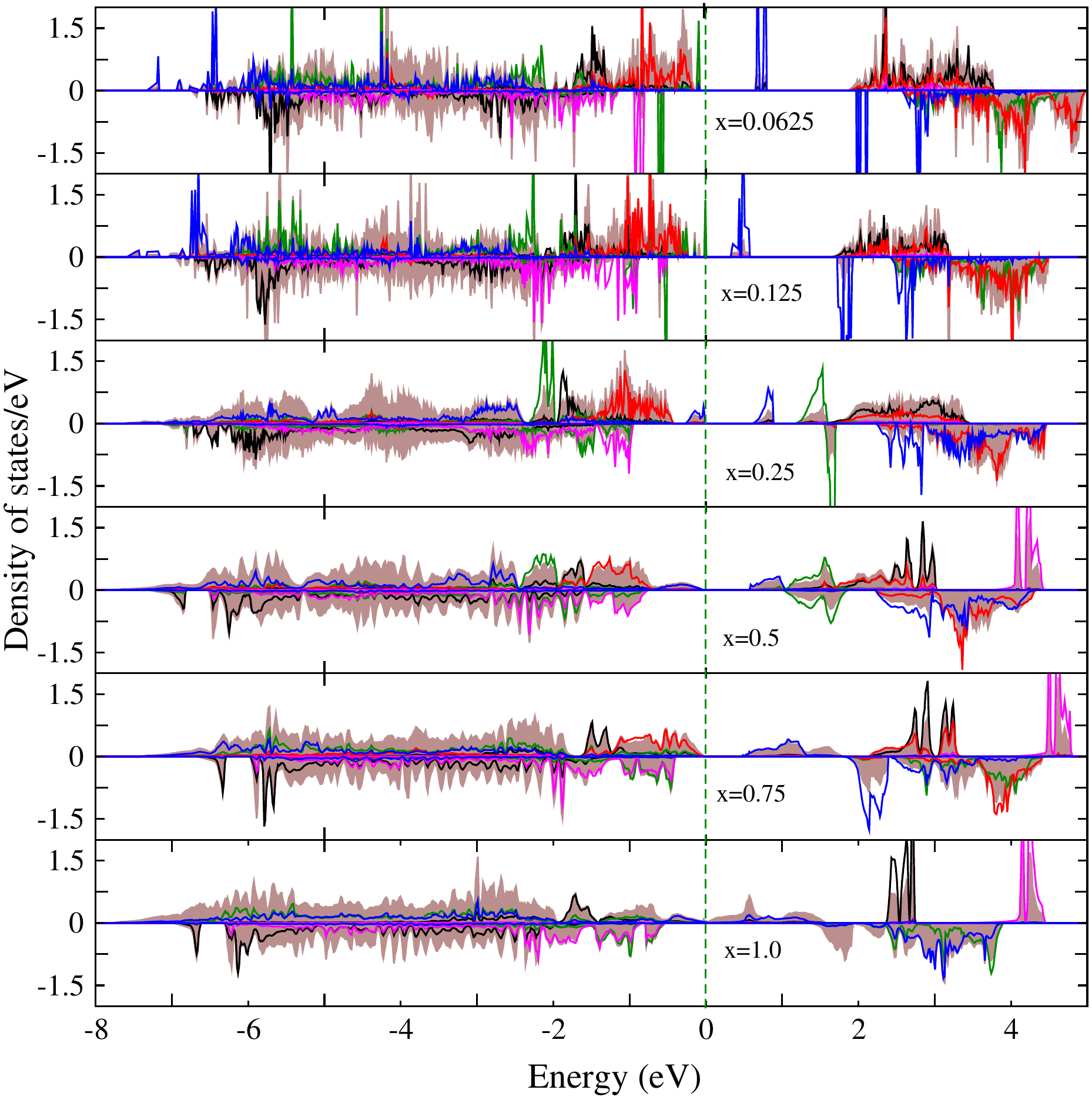}
\caption{\label{fig4} Total and atom-projected densities of states of 
$Co\left(Cr_{1-x}Mn_{x} \right)_{2}O_{4}$
for different $x$. The results are for $50 \%$ cation disorder ($y=0.5$).Here          
the total densities of states are denoted by brown shades. The black and the 
green curves 
represent atom projected densities of states for Co at tetrahedral and 
at octahedral sites respectively, the red curve represents atom
projected densities of states for Cr, the purple and the blue curves represent atom
projected densities of states for Mn atoms at tetrahedral and at octahedral sites
respectively.}
\end{figure}
\vspace{5 em}

\begin{figure}[ht]
\includegraphics[width=8.0cm,height=10.0cm]{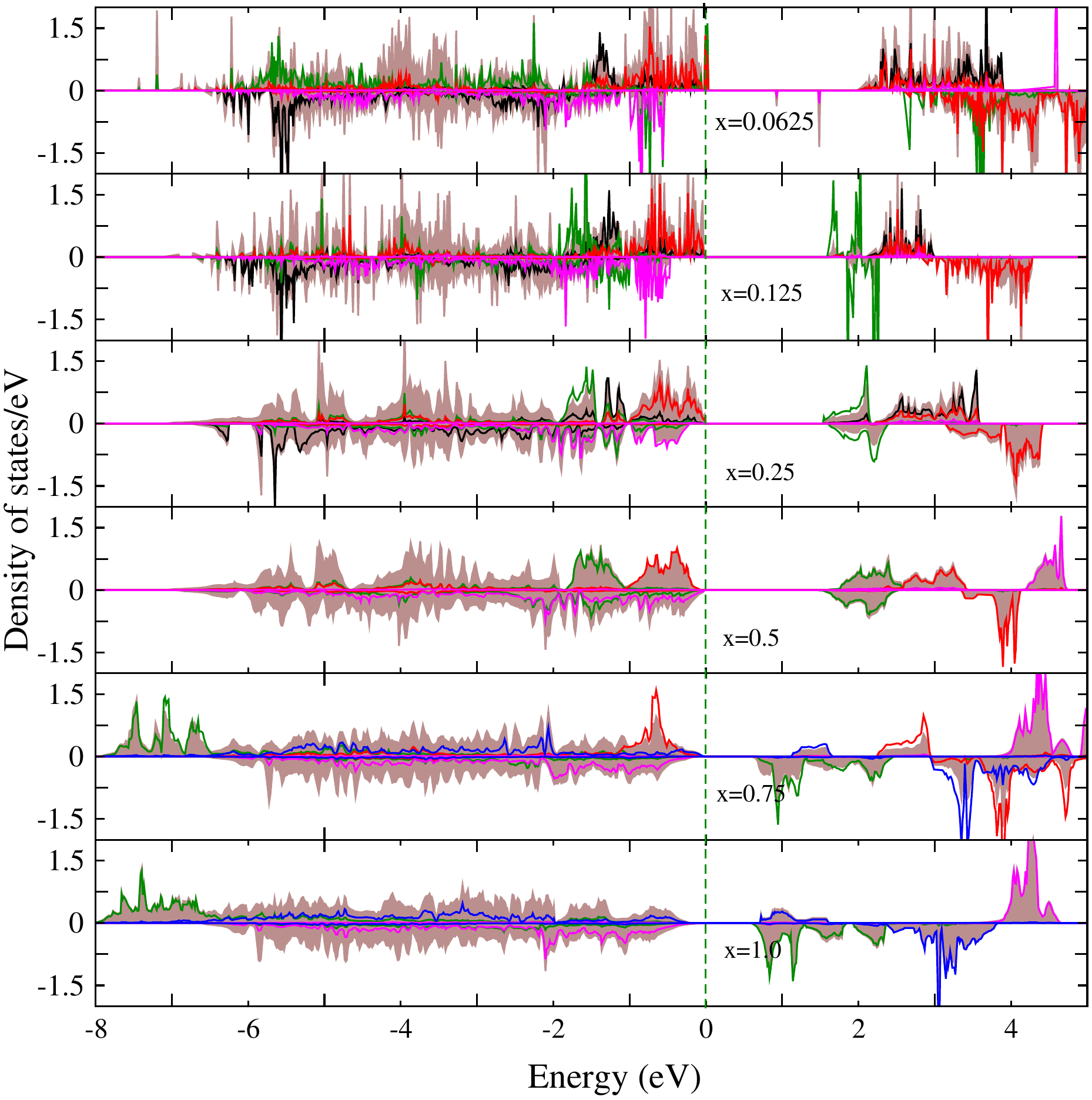}
\caption{\label{fig5}Total and atom-projected densities of states of                
$Co\left(Cr_{1-x}Mn_{x} \right)_{2}O_{4}$
for different $x$. The results are for full inverse arrangement ($y=1$).Here
 the total densities of states are denoted by brown shades. The black and the 
green curves
represent atom projected densities of states for Co at tetrahedral and 
at octahedral sites respectively, the red curve represents atom
projected densities of states for Cr, the purple and the blue curves represent atom
projected densities of states for Mn atoms at tetrahedral and at octahedral sites
respectively.}  
 
\end{figure}
Finally we present results on the electronic structures with variations in $x$ for three
degrees of cation disorder represented by $y=0,0.5$ and $1$ in Figures 
\ref{fig3},
\ref{fig4},and \ref{fig5} respectively. For the "normal spinel" configuration ($y=0$), the significant
changes with $Mn$ content happens in the majority (spin up) band, where $Mn$
states occupy the states near Fermi level moving the $Cr$ states to lower energies.
At low concentrations of $Mn$, sharp peaks corresponding to half-filled $e_{g}$
states, are observed on both sides of Fermi level which smear with increase 
in the $Mn$ content. The $Mn$ and $Co$ states dominate the majority band with 
increasing $x$ as expected. The $Mn$ $t_{2g}$ contributions are around 3-4 eV below
Fermi level, the $Co$ $e_{g}$ states contribute mostly around 2-3 eV below Fermi level.
The minority band consists of $Co$ for energies below Fermi level. Thus, across the 
concentration range, $Co$ is in a +2 state with fully filled $e_{g}$ and half-filled $t_{2g}$
orbitals, $Cr$ is in a +3 state with half-filled $t_{2g}$ and empty $e_{g}$ orbitals and
$Mn$ is in a +3 state with completely filled $t_{2g}$ and half-filled $e_{g}$ orbitals. 
Each of the three atoms retain their characteristics across the concentration range with 
the peaks corresponding to each of them occurring in distinctly separate energy regions.
Due to the appearance of $Mn$ states in the gap after Fermi level, the band gap decreases
with $x$ upto $0.5$ as reported in Table \ref{table5}. For $x=0.75$ and $1$, the band gap
again increases due to changes in the characteristics of the densities of states with the
structural distortions. This is due to the shift of the $Mn$ states near Fermi level and in the
occupied part towards lower energy, At these concentrations, the majority and minority
bands localise more, opening a larger gap. However, in going from $CoCr_{2}O_{4}$
to $CoMn_{2}O_{4}$, the band gap reduces considerably, due to the more delocalised
$Mn$ states as is apparent from the electronic structures presented here.

With $50 \%$ cation disorder($y=0.5$), new and interesting features appear 
in the electronic structures. Like $y=0$, we have sharp $e_{g}$ peaks coming from
octahedral $Mn$ atoms in the unoccupied part of the majority band close to the Fermi
energy for $x=0.0625$ and $0.125$. The contributions close to Fermi energy in the
occupied part of the majority band now comes from $e_{g}$ states of 
$Co$ at the octahedral sites. The octahedral $Mn$ contribution in the unoccupied part 
moves towards the Fermi level as $x$ changes from $0.0625$ to $0.125$, thus 
reducing the band gap, as happened for $y=0$ case. For these concentrations, the 
minority states near the Fermi level are made up of $t_{2g}$ orbitals of octahedral $Co$ 
and $t_{2g}$ states of tetrahedral $Mn$ atoms. At these concentrations, both octahedral
and tetrahedral $Co$ atom densities of states show completely filled $e_{g}$ and 
half-filled $t_{2g}$ orbitals. Thus $Co$ at both sub-lattices are in +2 charge state.
The densities of states significantly change at $x=0.25$ and continue to be so at $x=0.5$.
Now, the majority states in the occupied and unoccupied parts closest to the Fermi level
are again due to the octahedral $Mn$ atoms which increases the band gap in comparison to
that at $x=0.125$. This happens as octahedral $Co$ states in the majority band are 
sharply localised accommodating only the $t_{2g}$ electrons. The $e_{g}$ states are now
in the unoccupied part of the majority band and lie around 1-1.5 eV above Fermi level.
The minority octahedral $Co$ states consist of $t_{2g}$ in the occupied part and 
$e_{g}$ in the unoccupied part. Thus, the $t_{2g}$ orbitals of octahedral $Co$ are 
completely full and the $e_{g}$ orbitals are completely empty. The octahedral $Co$ atoms,
therefore, are in low spin states and in a +3 charge configuration. This is consistent with 
our results on magnetic moments and the explanations based upon bond lengths. The 
strong splitting and localisations of the $e_{g}$ and $t_{2g}$ states also signify a 
stronger crystal field. The features in the electronic structures, thus, support our arguments
based upon crystal field theory given in earlier sub-sections. As a consequence,
the band gaps increase from it's value at $x=0.125$. The densities of states
again change substantially with $x=0.75$ onwards. The localised octahedral $Co$ 
$t_{2g}$ states for $x=0.25,0.5$ now delocalise considerably. The contributions due to 
$e_{g}$ states of octahedral $Co$ appear at lower energies (at about -6 eV) indicating
reduction of the crystal field and return of octahedral $Co$ to a +2 charge state. For these
concentrations octahedral $Mn$ start occupying the majority states near Fermi level and 
with a larger structural distortion at $x=1$ in comparison to $x=0.75$, the octahedral $Mn$
$e_{g}$ states get energetically closer, reducing the band gap. Thus the band gap reduces to
zero at $x=1$. The highlights in the densities of states of the other component 
in the cation disorder, the $Mn$ atoms at the tetrahedral site have completely filled $t_{2g}$
and $e_{g}$ states in the minority spin channel along with 
completely empty majority spin channel. This implies that the electronic configuration of
this $Mn$ would be $\left(e_{g}^{2}t_{2g}^{3} \right)$, a +2 charge state. Once again the
results of the electronic structure confirms our picture on charge states as was discussed
in earlier sub-sections.  

With "complete inversion" that is for $y=1$, we now see larger splittings in the majority 
bands for any given concentration, in comparison to $y=0,0.5$. Now the octahedral
sites consist of only $Cr$ and $Co$ atoms upto $x=0.5$. In Reference \cite{dd2}, 
the octahedral crystal field parameters of $Cr$ and $Mn$ were calculated from
first-principles. These were extracted from the pristine $CoCr_{2}O_{4}$ and 
$CoMn_{2}O_{4}$. The results showed that the crystal field of $Cr$ is much stronger
than that of $Mn$. In here, we have seen that the states on either side of the Fermi
level are occupied by the octahedral atoms. Thus, when the octahedral sites are 
occupied by primarily $Cr$, it's strong crystal field makes the band gaps larger in
comparison to those for other degrees of cation disorder. With increasing $Cr$
substitution, the crystal field at the octahedral site starts losing it's strength, a signature 
of which is in the appearances of octahedral $Co$ $e_{g}$ states between 1-2 eV above
Fermi level. Consequently, the band gap decreases with increase in $Mn$ concentration.
A significant reduction in the band gap is observed when $x$ changes from $0.5$ to
$0.75$. In this case, the $Mn$ atoms have started to occupy the octahedral sites
which reduces the crystal field further. However, another interesting feature of the
densities of states at this concentration is that the tetrahedral $Co$ states in the 
majority bands are now localised at a significantly lower energy region. Concurrently,
there are more states of tetrahedral $Co$ in the unoccupied part of the minority band
at energies closer to the Fermi level. This is the reason behind drastic reduction in the
band gap. This happens due to the alteration in the charge state of octahedral $Co$.
Upto $x=0.5$, this $Co$ was in a low spin state due to the strong crystal field which 
paired all $t_{2g}$ electrons leaving $e_{g}$ states empty. A  inspection of the 
densities of states reveal that the crystal field is still strong with a significant 
splitting of $e_{g}$ and $t_{2g}$ states. Moreover, a comparison of octahedral
$Co$ densities of states for $y=0.5$ and $y=1$ for this concentration show that 
in case of $y=0.5$, the minority band had more densities of states in the energy
region -0.5 to -2 eV below Fermi level. This implies that for $y=1$, the octahedral 
$Co$ has less electrons, and thus the charge state is probably +3. In the high spin
state, octahedral $Co^{3+}$ has a higher OSPE \cite{cf}than $Co^{2+}$. Thus
when all $Co$ are made to occupy the octahedral sites as is done in case of $y=1$,
the expected charge state is +3. Therefore, the inferences from the features of the densities
of states and the crystal field results are consistent. The tetrahedral $Mn$ densities 
of states are by and large similar to that for $y=0.5$ signifying that it's charge state
is insensitive to the degree of cation disorder.

One outcome of the systematic explorations of the densities of states is the variations
in the nature of the band gap. Since the band gap depends both on the concentration
and degree of cation disorder quite substantially, manipulation of the composition and
the cation disorder can be a route to engineering band gap in this material which can be
subsequently used for applications like solar cells. This widens the functional scope of this 
material.

\section{Summary and Conclusions}
With Density functional theory based techniques, we have investigated the 
thermodynamics of cation disorder, the structural, the magnetic and the electronic
properties of $Co\left(Cr_{1-x}Mn_{x} \right)_{2}O_{4}$ compounds and analysed the
results from their electronic structures and the elements of the crystal field theory. 
Our results support the model of cation disorder between $Co$ and $Mn$ atoms 
as proposed by the experimentalists.
The experimental non-regular behaviour of magnetisation as a function of $Mn$ concentration
is explained on the basis of that.
By generalising a thermodynamic model of cation disorder for $AB_{2}O_{4}$ compounds
in this case, in conjunction with first-principles total energy calculations, we have quantified
the cation disorder parameter for each concentration. Our results have explored the intimate
relationships between the degree of cation disorder, the crystal fields associated with 
different atoms and their charge states, the structural and the magnetic properties. We have
demonstrated that the non-regular behaviour of the magnetic moments and the structural
properties can be traced back to the features in their electronic structures. Our results
show that the occupancies at the octahedral sites and the associated crystal fields can
explain the variations in the properties with varying composition and degree of cation 
disorder. By calculating the variations in the electronic band gap with variations in
composition and degree of cation disorder, we have shown that the functionalities of this
material can be enhanced by engineering the band gap through careful manipulation of the
composition and the degree of disorder. 

Overall, this work has paved a way to compute the thermodynamics of cation disorder 
in a $A\left(B_{1-x}C_{x} \right)_{2}O_{4}$ magnetic spinel and perform subsequent 
analysis to understand the microscopic details of such systems. In the context of the
specific system considered in this work, our results have provided a robust theoretical
background for interpretation of the experimental results.

\section{Acknowledgments}
The computation facilities from C-DAC, Pune, India
and from Department of Physics, IIT Guwahati funded under the FIST programme of DST, India
are acknowledged.

\end{document}